\documentclass[conference]{IEEEtran}
\IEEEoverridecommandlockouts
\usepackage{cite}
\usepackage{amsmath,amssymb,amsfonts}
\usepackage{algorithmic}
\usepackage{graphicx}
\usepackage{textcomp}
\usepackage[caption=false]{subfig}
\usepackage{xcolor}
\usepackage{microtype}        
\usepackage{float}    
\usepackage{placeins} 
\usepackage{comment}
\usepackage{hyperref}
\usepackage{enumitem}
\usepackage{cite}
\usepackage{tabularx}
\usepackage{placeins}
\usepackage[compatibility=false]{caption}
\usepackage{booktabs}
\usepackage{float}

\usepackage[none]{hyphenat}

\usepackage{fancyhdr}
\begin{document}
\raggedbottom
\setlist[itemize,1]{leftmargin=\dimexpr 25pt \relax}   
\setlist[itemize,2]{leftmargin=\dimexpr 25pt \relax}   
\setlist[itemize,3]{leftmargin=\dimexpr 25pt \relax}   

\title{Dynamic Evidence Collection Ecosystem for Assessment Integrity and Authentic Competence}

\author{

\author{\IEEEauthorblockN{Rajan Kadel}
\IEEEauthorblockA{\textit{NAPS,} 
Melbourne, Australia\\
Rajan.Kadel@naps.edu.au}
\and
\IEEEauthorblockN{Bellal Hossain}
\IEEEauthorblockA{\textit{NAPS,} 
Sydney, Australia\\
Bellal.Hossain@naps.edu.au}
\and
\IEEEauthorblockN{Samar Shailendra}
\IEEEauthorblockA{\textit{MIT,}
Melbourne, Australia \\
SShailendra@mit.edu.au}
\and
\IEEEauthorblockN{Bushra Naeem}
\IEEEauthorblockA{\textit{NAPS,} 
Sydney, Australia\\
Bushra.Naeem@naps.edu.au}
}}
\maketitle
\thispagestyle{fancy} 

\IEEEpubid{\begin{minipage}{\textwidth}\ \\ \\ \\ \\ \\ [10pt]
\centering
\end{minipage}}
\begin{abstract}
Generative Artificial Intelligence (GenAI) can produce high-quality essays, code, and design artefacts, challenging the validity of conventional assessments that rely on single-point submissions and product-only grading. This paper proposes a design framework called \textit{``Dynamic Evidence Collection Ecosystem''} that shifts assessment toward continuous, authentic, multi-source evidence of student learning over time. The framework collects process evidence through iterative artefacts, design logs, activity rounds, self-reflection, and peer collaboration, supported by an AI-enabled layer for learning analytics, formative feedback, and transparency. The approach is grounded in recent assessment-redesign scholarship in AI-rich contexts and aligned with contemporary views of authenticity in assessment.
This paper builds on the hypothesis that academic integrity is strengthened when it is treated as an assessment design rather than as an AI detection problem. The tools have limitations and risks of use that carry academic penalties. 
This paper presents an implementation scenario to support institutional adoption. 
\end{abstract}


\section{Introduction}
Generative Artificial Intelligence (GenAI) has accelerated long-standing tensions in educational assessment: the gap between what assessments intend to measure and the evidence they actually capture. As GenAI systems increasingly generate essays, code, reports, and design outputs, single-artifact submissions have become less reliable indicators of students’ competence. This has prompted calls to redesign assessment so that integrity becomes a feature of design. Recent work shows that educators are actively motivated to redesign assessments to maintain integrity, prepare learners for future work, and align with institutional policy and technological reality \cite{khlaif2025redesigning}.

Although the sector’s first reaction was to implement AI‑detection tools, their use is undermined by substantial reliability issues and ethical risks. Research evaluating AI-content detection tools reports inconsistent performance and high false positives, especially as model outputs improve and hybrid human–AI writing becomes more common \cite{sun2026trusting,weber2023testing,hadra2026evaluating}. Due to risks of false accusations and non‑transparent methods, several universities have disabled AI‑detection tools \cite{Curtin2026GenAIUse,WaterlooTurnitinAIDetection2025}.



Key issues for the assessment redesign are:
\begin{itemize} [leftmargin=4mm]
     \item Large Language Models make it easy to produce ``final product", creating a gap between correct submissions and genuine mastery of the underlying skills.
    \item This gap raises a key issue: how can we make the learning process visible to uphold academic integrity?
\end{itemize}

This article makes the following contribution:
\begin{itemize} [leftmargin=4mm]
     \item It proposes a framework for a \textit{Dynamic Evidence Collection Ecosystem}, positioning it as a process‑oriented response to GenAI‑driven pressures on assessment, and
     \item It introduces ``3-2-1 Portfolio Assessment" as an example for assessment design.  
\end{itemize}

The paper is organised as follows: Section~\ref{Sec:RW} reviews relevant literature and identifies the gap; Section~\ref{sec:framework} presents the proposed framework; Section~\ref{sec:Design} demonstrates its application; and Section~\ref{Sec:Conclusion} concludes the study.



\IEEEpubidadjcol
\section{Related Works} \label{Sec:RW} 

GenAI has significantly challenged the core principles underlying traditional assessment practices in higher education \cite{cotton2024chatting, xia2024scoping}. Conventional assessment methods have primarily emphasised product-based evaluation, typically in the form of a single final submission such as an essay, report, or examination, a model that has long been considered a cornerstone of measuring student learning \cite{xia2024scoping, chiu2024future}. The worth of such products has been significantly diminished by the capacity of GenAI tools to assist students in completing tasks such as essay writing, proposal writing, and take-home examinations, raising serious concerns about increased risks of cheating and the undermining of academic integrity \cite{belkina2025implementing, xia2024scoping}. Therefore, students are increasingly able to bypass the learning process by submitting AI-generated work as their own, gaining an unfair advantage over peers \cite{luo2024critical}. Acknowledging this concern, Moorhouse et al. found, through a review of assessment guidelines from the world's top-ranked universities, that the redesign of assessment is increasingly being encouraged, with a shift toward emphasising critical thinking and focusing on the learning process rather than on final outputs \cite{moorhouse2023generative}. Furthermore, the
authors in \cite{francis2025generative} recommend that institutions develop GenAI literacy initiatives, redesign assessments to prioritise critical thinking and creativity, and implement transparent, equitable policies governing GenAI use. Taken together, this body of evidence underscores that the conventional product-focused, single-submission model of assessment is no longer adequate in the GenAI era and requires fundamental redesign \cite{moorhouse2023generative, francis2025generative, crawford2023leadership}.

In response to the threats GenAI poses to academic integrity, many higher education institutions have turned to AI detection tools, such as Turnitin AI, GPTZero, and ZeroGPT, in an attempt to identify AI-generated content in student submissions. However, a growing body of peer-reviewed evidence demonstrates that these tools are fundamentally unreliable and unsuitable as a primary basis for determining disciplinary outcomes \cite{weber2023testing, dalalah2023false}. Statistical experiments of testing the rates of false positive and false negative detection of AI-generated text, concluding that false positives represent a particularly serious concern and the current detection tools need major improvements before they can be reliably used in academic contexts \cite{dalalah2023false}. Moreover, current AI detection tools are inaccurate and unreliable, often misclassifying AI-generated text as human-written, as reported in \cite{weber2023testing}. Furthermore, commercial tools that ``humanise" AI-generated text make it easy to bypass detection, meaning weakness in detectors is quickly exploited, and the systems become even less reliable \cite{ardito2025generative}. Alongside these technical challenges, AI content detectors often misclassify non-native English writing as AI-generated, raising significant concerns about fairness and the potential marginalisation of non-native English speakers in assessment and educational contexts \cite{liang2023gpt}. Scholars broadly advocate for a fundamental redesign of assessment as a more pedagogically sound, equitable, and sustainable response to the challenges posed by GenAI \cite{moorhouse2023generative, dalalah2023false}. Therefore, there is an urgent need for assessment redesign in the era of GenAI.

\section{Dynamic Evidence Collection Ecosystem}\label{sec:framework}
\begin{figure}
    \centering
    \includegraphics[width=1\linewidth, keepaspectratio]{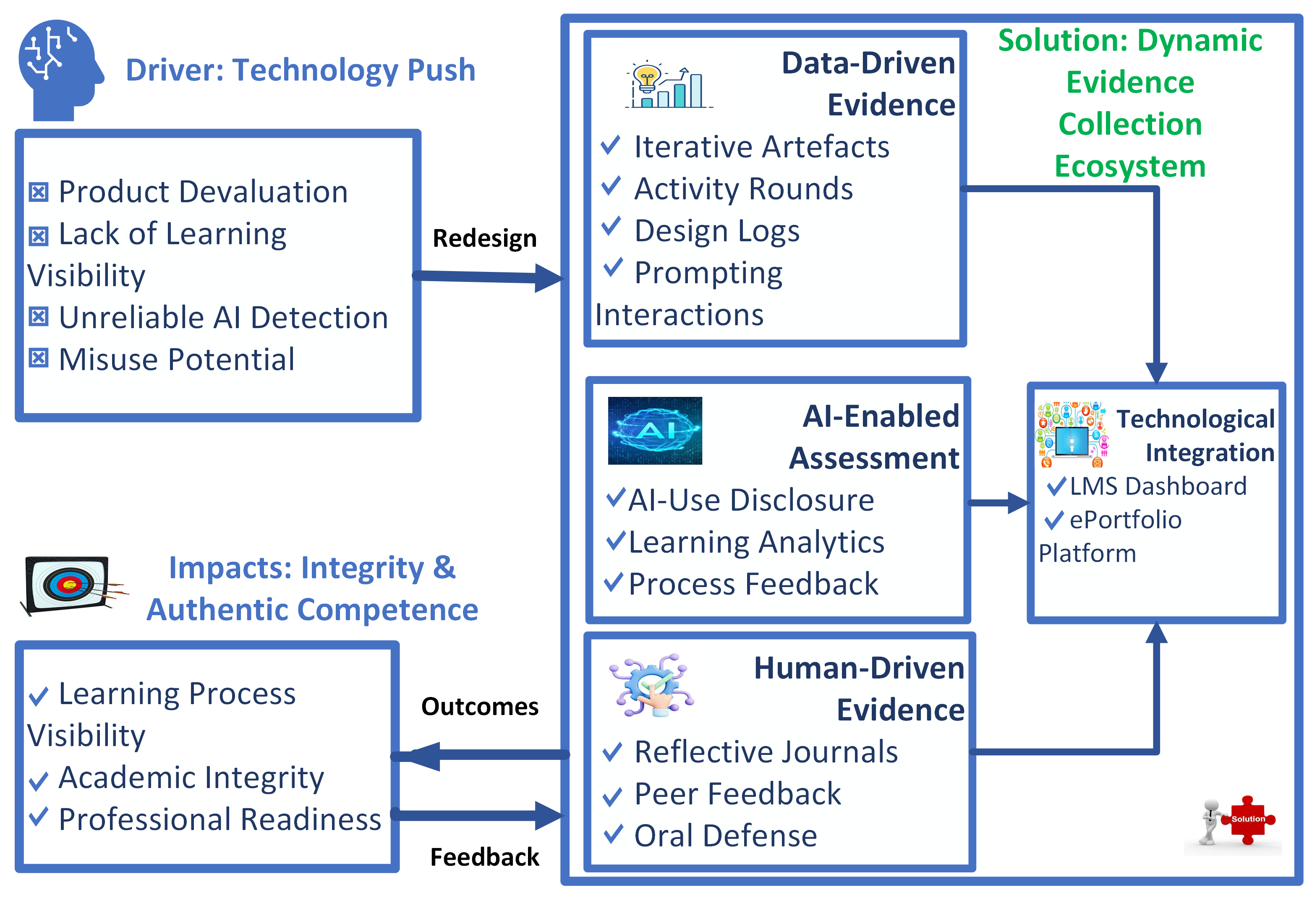}
    \caption{Framework for Dynamic Evidence Collection Ecosystem}
    \label{fig:model1}
\end{figure}

The literature review argues that traditional, single-submission assessments are increasingly vulnerable to AI misuse and a lack of visibility into the learning process. To address these concerns, we introduce a \textit{Dynamic Evidence Collection Ecosystem} as a more authentic, robust alternative that supports learning visibility, academic integrity and professional readiness. Fig.~\ref{fig:model1} presents a three-component framework for authentic assessment.

\subsection{Driver: Technology Push} 
This component highlights why we need to do assessment redesign in the era of GenAI. The key driving factors are:
\begin{itemize} [leftmargin=4mm]
    \item \textit{Devaluation of Product:} AI can generate an essay, report, design or code, the ``single artefact" loses its validity as proof of learning. Therefore, there is a risk that students may misuse it. Moreover, AI detection tools are unreliable which suggests that  policing AI use is neither scalable nor pedagogically sound \cite{elkhatat2023evaluating,sadasivan2023can}.
    \item \textit{Hidden Process Problem}: Traditional assessment sees outcome but not the cognitive process, making it impossible to distinguish between student effort and AI output \cite{zhan2025designing,ajjawi2024authentic}.
\end{itemize}

\begin{table*}[t]
\centering
\footnotesize
\caption{Mapping Evidences to Their Purposes, Applications, Examples, and Measured Skills.}
\label{table:mapping} 
\setlength{\tabcolsep}{3pt} 
\renewcommand{\arraystretch}{1.0} 
\begin{tabularx}{\textwidth}{@{}p{1.5cm} p{5cm} p{2.8cm} p{3.4cm} X@{}}
\toprule
\textbf{Evidence} & \textbf{Purpose} & \textbf{Applications} & \textbf{Examples} & \textbf{Measured Skill(s)} \\
\midrule

Iterative design artefacts & Captures the evolution of design thinking, documenting iterative refinement, informed decision-making, and structured problem-solving &
Product development, software engineering, and complex problem-solving &
Computer-Aided Design (CAD) models; network designs; circuit schematics; technical drawings &
Complex systems design; Iterative solution development and refinement; Decision making; Design thinking; Problem solving \\ \hline

Version controlled records & Provides a transparent and traceable account of the development process: individual contributions, decision-making, and changes &
Computer science, software engineering, and data science &
Code commits; version histories; debugging logs; feature iterations &
Proficiency in engineering tools; individual contribution; systematic development practices \\ \hline

Simulation \& modelling outputs & Demonstrates analytical reasoning through structured experimentation, evaluation, and validation of solutions &
Engineering, physics, networking, and AI/ML domains &
Simulation results; parameter tuning records; performance evaluations &
Problem analysis; solution testing and evaluation; analytical thinking; modelling proficiency \\ \hline

Experiment logs & Captures scientific reasoning through documentation of experimental design, observations, and iterative investigation &
Physics, chemistry, biology, and engineering laboratory contexts &
Experimental setups; recorded observations; measurement data; troubleshooting records &
Scientific inquiry; methodical experimentation; experimental proficiency; data analysis skills \\ \hline

Design decision logs & Demonstrates engineering judgement through structured documentation of decisions, trade-offs, and rationale &
Systems engineering, network design, and AI system development &
Architectural decisions; algorithm selection; trade-off analyses &
Decision-making; justification of technical trade-offs; engineering judgement; critical thinking \\ \hline

Prototype development & Demonstrates the application of theoretical knowledge through the design, construction, and iterative refinement of practical solutions &
Robotics, product design, and software engineering contexts &
Hardware prototypes; software prototypes; staged system implementations &
Implementation proficiency; practical engineering skills; knowledge application;\\ \hline

Project milestones & Enables continuous monitoring of learning progression through structured checkpoints and staged deliverables &
Capstone projects and engineering design projects &
Project proposals; progress reports; testing reports; final deliverables &
Project planning; engineering management; project management skills; development trajectory \\ \hline

Digital portfolio & Provides a comprehensive and integrated record of competence development, showcasing skills and learning progression &
Any program featuring project-based or experiential learning &
Collections of reports, design artefacts, simulations, and reflective entries &
Demonstration of integrated professional competence; support for program-level competency \\  \hline

Reflective learning journals & Supports metacognitive development through structured reflection on learning processes, decision-making, and progression over time &
Problem-based learning tasks; capstone projects; collaborative group work &
Regular reflective entries; documented problem-solving strategies; use of GenAI tools &
Lifelong learning; self-assessment; professional responsibility; continuous professional growth \\ \hline

Peer collaboration & Evaluates teamwork and collaborative effectiveness through documented interactions, contributions, and engagement &
Team-based projects in engineering, computing, and science disciplines &
Peer feedback, contribution statements, team meeting records &
Teamwork; communication skills; collaborative effectiveness; individual contribution \\
\bottomrule
\end{tabularx}
\end{table*}

\subsection{Solution: Dynamic Evidence Collection Ecosystem} 
The proposed ecosystem leverages multimodal data streams, utilises AI-enabled assessment strategies, and is seamlessly integrated within the Learning Management System (LMS). To achieve this multimodality, the framework categorises evidence into two streams: 
\subsubsection{Data-Driven Evidence} 
To restore assessment validity, the framework utilises a multi-layered collection of data-driven inputs, ranging from version-controlled technical artefacts to real-time learning analytics, ensuring a verifiable trail of student authorship, which tracks technical progression. Some examples of methods that can be used for the data-driven evidence are:
\begin{itemize}  [leftmargin=4mm]
    \item Iterative Artefacts: Version-controlled submissions (e.g., draft-versions, prototype iterations, feedback cycles, GitHub commits) that demonstrate incremental development of the product that captures the process. These iterative artefacts may be more suitable for designing and developing the product, but can be used for report writing.  
    \item Activity Rounds: Time-bound, repetitive cycles of task execution that provide longitudinal visibility. For example, concept proposal, draft development feedback, revision and improvement, and final artefact. The activity rounds may be suitable for problem-based or project-based tasks. 
    \item Design Logs: Granular documentation of technical decisions and problem-solving steps. For example, design decisions and rationale, problems encountered and solutions, iterations and design changes, and analysis of tools and resources used. This evidence may be more suitable for design tasks. 
    \item Prompting Interactions: Documented histories of student-AI dialogue to ensure transparency in AI-assisted workflows. Students distinguish their contributions from those of AI. 
\end{itemize}
\subsubsection{Human-Driven Evidence}
Complementing the data-driven evidence, human-driven evidence provides the human-in-the-loop verification necessary to confirm conceptual mastery presented in the product \cite{fajardo2025human}.
\begin{itemize}  [leftmargin=4mm]
    \item Reflective Journal: Metacognitive accounts where students justify their design choices and learning trajectory while developing the products (design, report, project, product, GenAI usage and any other outcomes).  
    \item Peer Feedback (Scaffolded with GenAI): Evidence of social learning and interaction with GenAI tools. Evaluate own and peer contributions within a team-based environment. 
    \item Oral Defence: Direct, real-time verification of authorship and conceptual understanding through spoken inquiry.
\end{itemize}
Balancing of those two types of evidence in the assessment is critical for learning process visibility. These two types of evidence should support each other, and educators should evaluate both pieces of evidence together.  Once gathered, these evidence streams are synthesised through a centralised technological hub, where AI acts as a facilitative partner.

\subsubsection{AI-Enabled Assessment}
The framework utilises AI tools to provide ``Process Feedback" and ``Learning Analytics" that support the assessment cycle. This ensures that AI enhances the educator’s ability to interpret student progress without compromising trust. This framework also captures students' AI-use disclosure \& attribution. An AI-enabled assessment strategy can be ensured by the following actions:
\begin{itemize}  [leftmargin=4mm]
    \item AI-Use Disclosure \& Attribution: Clear frameworks for students to document how, where, and why AI was utilised in their creative or technical process.
    \item Learning Analytics: Automated tracking of engagement patterns to identify students who is struggling or whose work patterns deviate significantly from their established norms.
    \item Process Feedback: AI-generated formative feedback provided during initial stages to guide students before final submission.
\end{itemize}

\subsubsection{Technological Integration}
To manage the high volume of multimodal data, the ecosystem requires a robust technological infrastructure that centralises evidence from various sources and provides the outcomes in a suitable form. This technological integration should support aggregating evidence using analytics; it tracks learning over time and supports feedback loops \cite{kannan2022facilitating,martinez2023content}. This process can be implemented by:
\begin{itemize} [leftmargin=4mm]
\item ePortfolio Platform: A longitudinal repository where students curate their iterative artefacts, reflective journals, and video records into a cohesive narrative of competence.
\item LMS Dashboard: A centralised visualisation tool for educators that aggregates data from the LMS, assessment works, design logs, and peer interactions to provide a ``heat-map" of class-wide and individual progress.
\end{itemize}

\begin{table*}[h!]
\centering
\footnotesize
\caption{3-2-1 Portfolio Assessment.}
\label{tab:321}
\begin{tabular}{|p{4.2cm}|p{11.5cm}|p{0.7cm}|}
\hline
\textbf{Component (Evidence Type)} & \textbf{Description (Timeline)} & \textbf{Weight} \\ \hline
Three (3) iterations (Data-driven) &  Versions (v1, v2, v3) with change notes and reasoning for the changes (Initial, Mid and Final)  & 40\% \\ \hline
Two (2) critiques (Human-driven) & Two reflections/peer-reviews that support iterations/changes (Mid and Post development) & 30\% \\ \hline
One (1) defence (Human-driven) &  Oral explanation, viva for defence that supports iterations and critique (Post development) & 30\% \\ \hline
\end{tabular}

\end{table*}

By triangulating these multimodal data streams, the ecosystem establishes a robust foundation for verifying authentic competence and maintaining academic integrity. 

\subsection{Outcome: Integrity and Authentic Competence} By strengthening evidence collection, the redesigned assessment improves learning visibility, academic integrity, and professional readiness. Educators can see how learning unfolds through continuous, contextual evidence that makes the learning process visible using the proposed framework. The ecosystem directly addresses hidden process concerns by providing continuous, observable traces of learning. This supports educators’ ability to make judgements about growth, decision-making, and engagement. 

It is far more difficult to fabricate a sustained, multi‑step process because integrity is built into a design that requires ongoing, authentic engagement. Academic integrity becomes an integral part of assessment design in this approach, as fabrication across both data-driven and Human-driven is considerably more difficult than generating a single polished product. This shifts integrity from reactive detection to proactive assessment architecture, consistent with literature emphasising process-based redesign for integrity in AI-rich environments.

The dynamic evidence collection ecosystem mirrors real-world practice, including product iteration, self-reflection, collaboration, and tool use, including AI usage. The ecosystem reflects professional work practices in many fields, where competence is demonstrated through iterative development, documentation, collaboration, and the responsible use of tools, which will enhance employability skills and job readiness, consequently, professional readiness in students.

Table~\ref{table:mapping} illustrates various evidence, their purposes, applications, examples and the mapping of measured skills achieved by evidence. Most of the evidence presented in the table is suitable for STEM education. However, a similar table can be created by educators in any discipline, utilising their expertise. Educators need to create a similar table, covering appropriate evidence for each subject, suited to the discipline. Then, educators need to develop a strategy to prepare assessments utilising the table. 


Educators reflect and evaluate on learning visibility \cite{hattie2008visible,guruge2023towards}, academic integrity and professional readiness \cite{orr2023systematic} impacted by the ecosystem using suitable techniques. The findings from the analysis are provided to the ecosystem as feedback for further improvement on evidence collection, strategies for using AI in assessment, and technological integration. Therefore, this ecosystem transforms assessment from static to dynamic and emphasises the process of observing learning visibility. 

\section{3-2-1 Portfolio Assessment}\label{sec:Design}
This section translates the ecosystem into an implementable assessment design. Educators replace “one-and-done” major assessment submission with a sequence of assessable artefacts, reducing dependence on any single product and strengthening validity through triangulation \cite{khlaif2025redesigning,ncube2026redefining,banihashem2025optimizing}. The assessment distributes evidentiary weight across time \& sources. Table~\ref{tab:321} shows one assessment example, called ``3-2-1 Portfolio Assessment", that shows components with types of evidence, a short description and possible weights. ``3" refers to three-iteration during product development to observe the development progress, ``2" refers to two critique documents covering reflections (self or group) and/or peer reviews completed during the product development, and ``1" refers to one clearly demonstrated defence. Educators can vary the portfolio according to the discipline, but there is a need to balance two evidence types to observe the learning visibility. 

A three-iteration scenario can encompass various types of assessments, and the assessment can be done in a group or individually. In this process, students documented changes, including how the formative feedback was addressed and the reasons for changes and decisions. Students need to declare their use of AI tools using AI-use disclosure \& attribution. 
The AI use statement should cover: a) What AI tool was used?, b) What are the purposes of the tools?, and c) What was changed by the student from the AI output? This process is critical for AI transparency. This supports authenticity in assessment by valuing contextual reasoning and evaluative judgement rather than just polished outputs. It also aligns with assessment redesign guidance that prioritises skills (ethical decision-making and trade-off reasoning), which are less easily replicated by GenAI. Sample examples for the three-iteration cases are provided for clarity:
\begin{itemize} [leftmargin=4mm]
    \item Design-based problem, such as system design, or software design requires students to maintain a decision ledger (chosen approach → options considered  → decision) for three stages. They are also required to maintain a design ledger to capture logs that record decisions, design constraints, and evidence. 
    \item Project-based or problem-based task where students are either working in a group or individually required to demonstrate progress at three checkpoints (scope, \& plan → prototype (design) → final product) and also address how the feedback is addressed from one checkpoint to another and document how changes are made.
    \item Report (essay) writing where students maintain their report writing versions (proposal → draft → final), where students record the changes. Students outline the reasons for the changes and address how the feedback is considered. 
\end{itemize}

Additionally, the portfolio includes two (2) critique documents (reflections and/or peer reviews, depending on the suitability and effectiveness) and one (1) oral defence work. Peer collaboration and reflections should support authenticity and practice-based learning and must be directly linked with the three-iteration stages. Therefore, critique documents must be structured and assessed to ensure that students understand how the quality of feedback and uptake of critique is handled, and how the group and individual progress. Reflection should be directly linked with the activities associated with the three iterations, the progress made and the plan. Structured peer evidence provides a practical mechanism for ensuring consistent, transparent, and reliable evaluation by guiding educators to comment on the same critical aspects of a work, thereby improving the quality and comparability of feedback. 

Finally, the oral defence provides a direct, real-time verification of authorship and conceptual understanding through spoken inquiry. It allows examiners to probe reasoning, clarify ambiguities, and assess the candidate’s ability to articulate and defend their work independently under questioning. The portfolio approach is consistent with systematic recommendations for multi-stage, process-based assessment to mitigate academic integrity risks in AI-infused environments. Educators need to evaluate the three components together by triangulating these multimodal data streams to establish a robust foundation for verifying competence. If there are any concerns about any evidence or missing components, there is a need for further investigation for academic integrity.

This work is conceptual and lacks empirical validation. Technological and governance requirements are not fully specified, which may affect scalability and adoption. Future work should address these limitations through pilot testing. 

 
 \section{Conclusion}\label{Sec:Conclusion}
GenAI poses significant challenges to the validity of assessment systems that rely on single‑point submissions and product‑focused grading. Emerging evidence demonstrates that AI‑detection technologies lack the reliability required for high‑stakes academic integrity decisions. These limitations underscore the need for assessment redesign that foregrounds learning visibility, supports integrity through process‑based evidence, and aligns more closely with contemporary professional expectations. This paper introduces a \textit{Dynamic Evidence Collection Ecosystem} as a practical design response to current assessment challenges. The proposed approach conceptualises evidence collection as a continuous, multi‑source record of learning, supported by formative analytics and strengthened through transparency norms. Realising this ecosystem requires a commitment to transparent governance and a streamlined evidence-based system to address concerns around a product-only approach, while ensuring that the increased visibility of the learning process does not result in prohibitive educator workloads. The ecosystem positions the assessment task as requiring both data‑driven and human‑driven forms of evidence to enhance process visibility, thereby promoting academic integrity and professional readiness. 
\bibliographystyle{IEEEtran}
\bibliography{Myref}

\vspace{12pt}

\end{document}